\documentclass[conference]{IEEEtran}
\IEEEoverridecommandlockouts
\usepackage{cite}
\usepackage{amsmath,amssymb,amsfonts}
\usepackage{algorithmic}
\usepackage{graphicx}
\usepackage{textcomp}
\usepackage{xcolor}
\def\BibTeX{{\rm B\kern-.05em{\sc i\kern-.025em b}\kern-.08em
    T\kern-.1667em\lower.7ex\hbox{E}\kern-.125emX}}
\begin{document}

\title{Revisiting the RBLE design based on Matlab simulation\\
}

\author{\IEEEauthorblockN{1\textsuperscript{st} Bohan Lou}
\IEEEauthorblockA{\textit{University of Science and Technology of China} \\
lbh2001@mail.ustc.edu.cn}
}

\maketitle

\begin{abstract}
As a key low-power communication technique, backscatter communication has received significant attention since the rising of the Internet of Things (IoT). We revisit the state-of-the-art backscatter system, RBLE [1]. It solves several key reliability issues of backscatter system including unreliable two-step modulation, productive-data dependency, and lack of interference countermeasures. We implement a Matlab simulation version of this. It uses the reverse whiten technique to generate a single tone signal, operates direct frequency on it and calculates the bit error rate (BER) to evaluate. We give the spectrograms of the middle waveform results, compare the influence of different modulation methods and analyze the cause of high BER. In the end, we discuss the future prospects of the applications using RBLE.
\end{abstract}


\section{Introduction}
Backscatter provides great potential for different Internet of Things (IoT) applications such as smart homes, supply chains and personal identification. It has the ability of harvesting energy from RF sources and providing ultra-low-power wireless communications. But past backscatter systems[2][3][4] need specialized and expensive receivers to decode the backscatter reflected signal. To solve that problem, there is a lot of research on designing backscatter systems that work with commodity radios.  Take WiFi backscatter as an example, WiFi backscatter system like [5][6] can decode the reflected signal by commercial WiFi receivers. 

RBLE is a robust BLE backscatter system that works with a single commodity receiver. Its idea comes from FreeRider .The whole system works completely on commodity Bluetooth radios. As shown in Figure 1, the signal source generates a single tone exciting BLE signal. Then it uses direct frequency shift modulation to regenerate a BLE packet that includes tag information and obeys the BLE signal format. It uses several techniques to enhance the robustness of the whole system. It introduces dynamic channel configuration that allows RBLE tags to perform channel hopping just like what Bluetooth Classic does. Also, during the BLE Packet Regeneration phase, it uses adaptive encoding to further enhance reliability in challenging situations.

\begin{figure}[ht]
    \centering
    \includegraphics[width=0.5\textwidth]{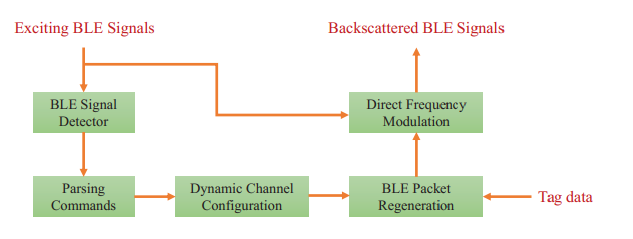}
    \caption{RBLE framework(This figure is from [1])}
    \label{fig:1}
\end{figure}

Matlab has mature toolbox for simulating BLE signals and it is convenient to operate direct frequency shift. This paper aims to review the structure and verify the results of RBLE by simulating it in Matlab. We focus on the BER given in the original paper. We rebuild all the components of RBLE except for channel hopping and the adaptive encoding part. 

As a result, some types of data can be transferred perfectly without any noise. But it behaves badly when adding noise to it. We analyze the reason in the evaluation part.

\section{Background}
\subsection{BLE}\label{AA}
BLE(Bluetooth Low Energy) is a low-power wireless technology used for connecting devices with each other. It was introduced in 2010 as part of the Bluetooth 4.0 specification. Compared to the standard Bluetooth protocol, BLE is targeted more towards applications that need to consume less power and run on batteries for longer periods of time. It focuses more on Internet of Things (IoT) applications where small amounts of data are transferred at lower speeds. For example, a growing number of hotels, school dormitories and department buildings are installed on the smart door lock. Most intelligent door locks use BLE Bluetooth technology so they don't need to replace the battery frequently.

\begin{figure}[ht]
    \centering
    \includegraphics[width=0.5\textwidth]{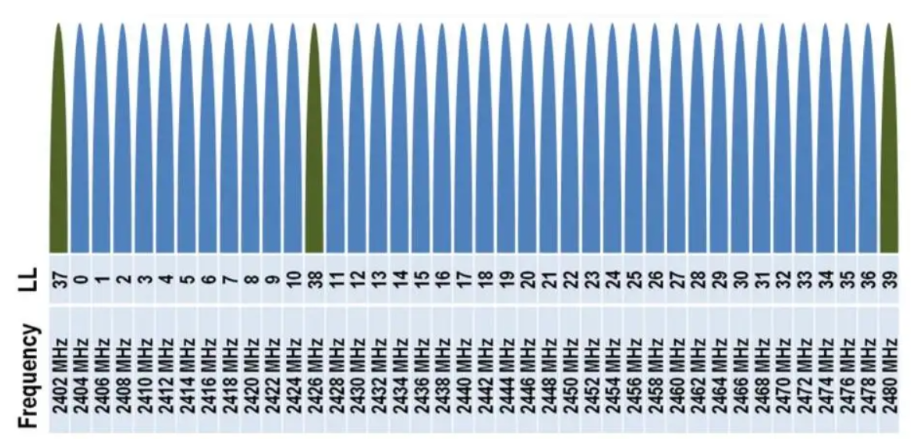}
    \caption{BLE Channel}
    \label{fig:14}
\end{figure}

BLE operates at the 2.4GHz ISM band which is the same spectrum used by Bluetooth Classic. As shown in figure 2, it has 40 pre-defined channels ranging from 2400 to 2483.5 MHz, each of which is 2 MHz wide. Within a channel, it uses Gaussian Frequency Shift Keying (GFSK) modulation to transfer data. It has a 1 Mbps raw bit rate and the maximum TX power is 10 mW.

\subsection{Backscatter}
Backscatter is a type of system that uses ambient backscatter signals for devices to communicate. It usually consists of three parts as shown in figure 3: the signal source, the backscatter transmitter equipped with antennas, and the backscatter receiver. The source signal is either from the environment or from a transmitter. The backscatter transmitter makes some changes to the signal to include its information. The backscatter receiver receives the signal and recovers the tag data. Different from traditional communications, which transmit data by generating an RF signal itself at the transmitter, backscatter communications convey information by remodulating and reflecting signals from other signal sources instead of generating carrier signal at its transmitter.

One of the main benefits of using backscatter is its low deployment cost. By using existing surrounding RF signals as the signal source instead of designing dedicated carrier emitters, the cost of ambient backscatter can be significantly lowered. This makes it an attractive option for applications in IoTs and green communications.

The most popular backscatter communication is the RFID [9] technology. RFID stands for Radio Frequency Identification. It is a technology that uses radio waves to transmit information wirelessly. RFID requires using a device known as a reader. The reader is needed to retrieve any data stored on an RFID tag. RFID technology is used for automatic identification and tracking of objects or entities by making use of radio-frequency waves. It is known to be a combination of radio broadcast technology and radar. In recent years, RFID technology has moved from obscurity into mainstream applications that help speed the handling of manufactured goods and materials.

\begin{figure}[ht]
    \centering
    \includegraphics[width=0.5\textwidth]{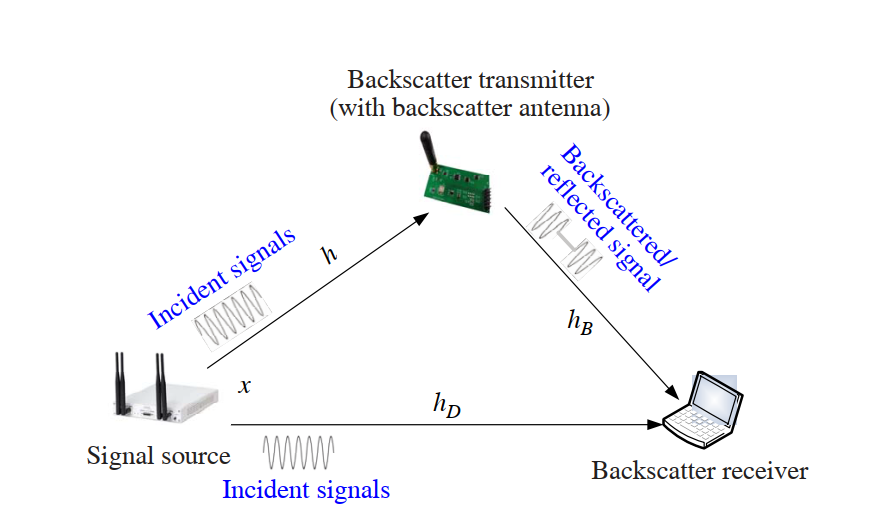}
    \caption{An example of signal transmission in backscatter communication system(This figure is from [8])}
    \label{fig:2}
\end{figure}

\section{RBLE simulation design}
\subsection{Overview}
Figure 4 shows the framework of the Matlab implementation of RBLE. A BLE device using the reverse whitening technique to generate a single tone exciting baseband signal. Then we convert it into a passband signal. We do the direct frequency shift bit by bit based on the tag data. Finally we convert the passband signal back to baseband and send it to a BLE waveform receiver to get the received data.

\begin{figure}[ht]
    \centering
    \includegraphics[width=0.5\textwidth]{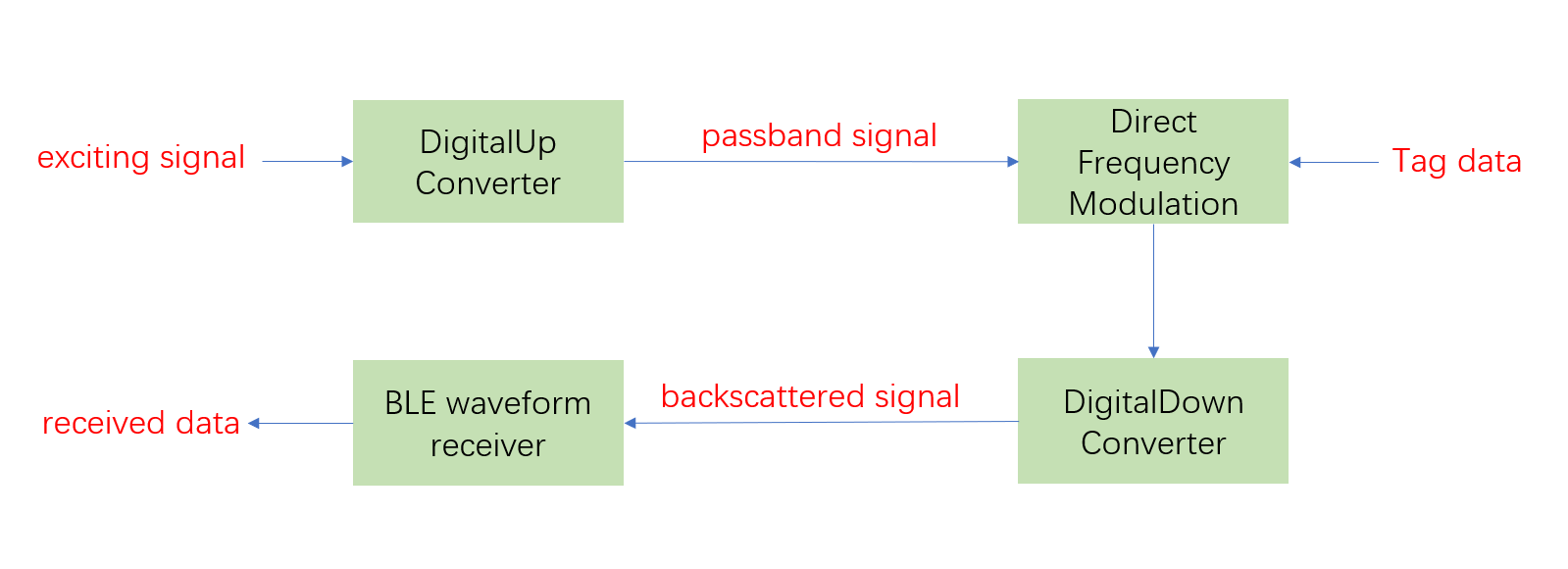}
    \caption{The data path of the RBLE Matlab simulation}
    \label{fig:3}
\end{figure}

\subsection{Reverse Whitening}
The exciting signal generated by the signal source should be single-tone in order to operate direct frequency shift on it. However, Bluetooth uses a technique called whitening to encode the message to keep the data from repeating 0s and 1s. Fortunately, the whitening algorithm is a symmetric encryption algorithm using channel ID as the initial whitening key. The whitening key is updated by the generator polynomial $g(x) = x9 + x^5 + 1$. The whitening algorithm will do an xor operation on the message and the whitening key. That implies that we can perform the same algorithm as the reverse whitening algorithm before transforming the data to the BLE waveform. By using this technique, we can make the payload part of BLE signals all zeros or ones, which are single tones.

\subsection{Digital Up Converter and Down Converter}
The signal that the Matlab BLE waveform generator produces is a baseband signal. We need to translate it from the baseband to the Intermediate Frequency (IF) band. We convert the baseband signal's frequency by multiplying it with a complex exponential with center frequency equal to the value in the CenterFrequency property. Translating it back to a baseband signal is the same. We use a cascade of three decimation filters to downsample the frequency.

\subsection{Direct Frequency Shift}
The core of RBLE is how to modulate an exciting BLE signal into another BLE signal. Before introducing the technique we use to modulate the tag data, let us have a short review of how original BLE signals modulate. In a channel of 2 MHz, the symbol 0 is represented by a negative frequency deviation of 250 kHz while the symbol 1 is represented by a positive frequency deviation of 250 kHz, following the Gaussian Frequency Shift Keying (GFSK). 
As mentioned in the reverse whitening part, the payload part of BLE signal can be made single tone. Next,
we directly apply frequency shifts to modulate 0 or 1 in the target channel. 

Here's an example in Figure 5. The all-zero single tone signal is on channel 37, where the center frequency is 2402 MHz. The target channel is channel 3, whose center frequency is 2410 MHz. When we want to add tag data 0, we do a frequency shift of 8 MHz to modulate. Similarly, we use 8.5 MHz as the shift frequency to modulate 1. The design is simple and easy to implement. As for the header part of the BLE signal, we reuse the original header to generate the new packet's header. Still the example of channel 37 and channel 3, we apply an 8 MHz frequency shift on the original header to get the new packet's header.

\begin{figure}[ht]
    \centering
    \includegraphics[width=0.5\textwidth]{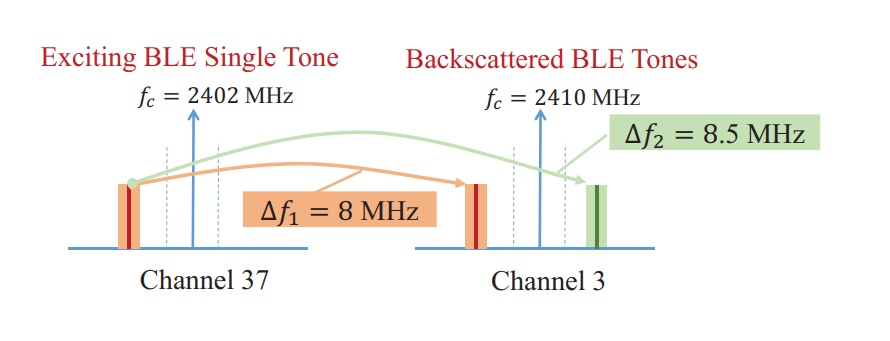}
    \caption{Direct frequency shift modulation example. To modulate a symbol
0 to target channel 3, we only need to shift a frequency of 8 MHz on the
zeros single tone of the exciting signal on channel 37. A frequency shift of
8.5 MHz is for the symbol 1.(This figure is from [1])}
    \label{fig:4}
\end{figure}

Note that the shift frequency is greater than the width of a BLE channel, the unwanted copy of our frequency shift will never fall into our target channel. It reduces the BER by a huge amount.

\subsection{AWGN}
Additive White Gaussian Noise (AWGN) is the most basic noise and interference model for simulating the channel between transmitter and receiver. The noise is added to the original signal. This model has constant spectral density and Gaussian amplitude distribution. We use that model because it is close to the real-world BLE signal transfer situation.

\section{implementation}
We build the RBLE simulation project based on Matlab R2020b and use its Communications Toolbox. The implementation is detailed as follows. 

\textbf{Basic waveform structure.} The BLE baseband waveform is a time-domain signal for the given input message bits. It is a complex-valued column vector of size Ns-by-1, where Ns represents the number of time-domain samples. 
\begin{equation}
Ns = sps * (HeaderLen + DataLen)
\end{equation}
As shown in Equation (1), sps represents symbols per sample. HeaderLen means the length of the header, which depends on the BLE specification. DataLen is the length of the data that we want to transmit. The structure of the BLE packet is shown in Figure 6. In that case, HeaderLen is 40.

\begin{figure}[ht]
    \centering
    \caption{Data structure of BLE packet in Matlab}
    \begin{tabular}{|c|c|c|}
        \hline
        preamble  & access address & data \\
        (8 bits)  & (32 bits) & ( $<2080$ bits) \\
        \hline
    \end{tabular}
\end{figure}

\textbf{BLE transceiver.} We use bleWaveformGenerator() as the BLE signal transmitter and bleIdealReceiver() as the BLE signal receiver. The default modulation method is Gaussian Filtered Minimum Shift Keying (GMSK). We implement another version of the transceiver using FSK modulation. GFSK is not implemented in our project because it will introduce extra delays. We choose the mode 'LE1M' so it transmits at 1 Mbps raw data rate. The frequency deviation is 250 kHz.

\textbf{RBLE tag.} We add the tag data by directly multiplying a cosine function to the IF band. As shown in the Equation (2), the variable t represents the time domain of the IF band and f represents the shift frequency. For the case in Figure 5, f is 8,000,000 for tag data 0 and 8,500,000 for tag data 1. 

\begin{equation}
TAGwaveform = IFwaveform * cos(2 \pi f t);
\end{equation}

\textbf{Metrics.} We use the bit error rate (BER) as the core performance metric. The low BER shows the reliability and robustness of the RBLE design. We change the type of tag data, noise intensity and modulation method for BLE signals to see the influence of these factors on BER. The noise intensity is controled by Eb/No. The relationship between Eb/No and the signal noise ratio (SNR) is stated by Equation (3).
\begin{equation}
SNR = EbNo - 10 * log10(sps)
\end{equation}

\section{Evaluation}
\subsection{Experiment Setup}
The exciting signal is on channel 37 with center frequency of 2402 MHz. The target channel is channel 3 with center frequency of 2410 MHz. The frequency deviation is 250 kHz. 

\subsection{Single Tone Baseband Waveform}
Figure 7 shows the amplitude-time graph of the single tone baseband waveform generated by the reverse whitening technique. It shows the correctness of a single tone waveform because the real and imaginary parts are constant trigonometric functions.

\begin{figure}[ht]
    \centering
    \includegraphics[width=0.5\textwidth]{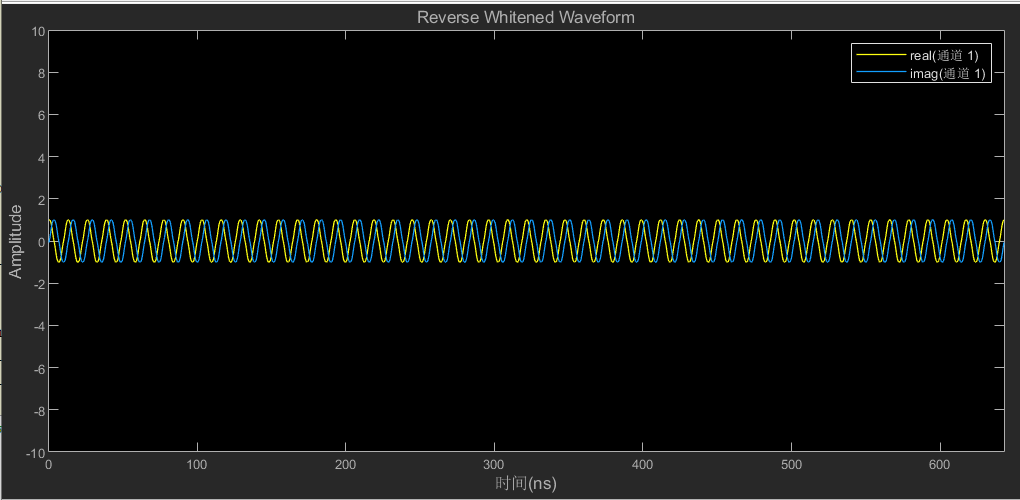}
    \caption{The amplitude-time graph of the single tone baseband waveform}
    \label{fig:6}
\end{figure}

\subsection{Original Waveform And Reflected Waveform}
Figures 8 and 9 are spectrograms of the exciting signal and the reflected signal. The center frequency of the exciting signal is 2402 MHz. The reflected signal has two peak frequencies of 2410 MHz and 2394 MHz. The unwanted copy of 2394 MHz is outside of BLE Channel 3. It shows that the direct shift frequency operation has been done successfully.

\begin{figure}[ht]
    \centering
    \includegraphics[width=0.5\textwidth]{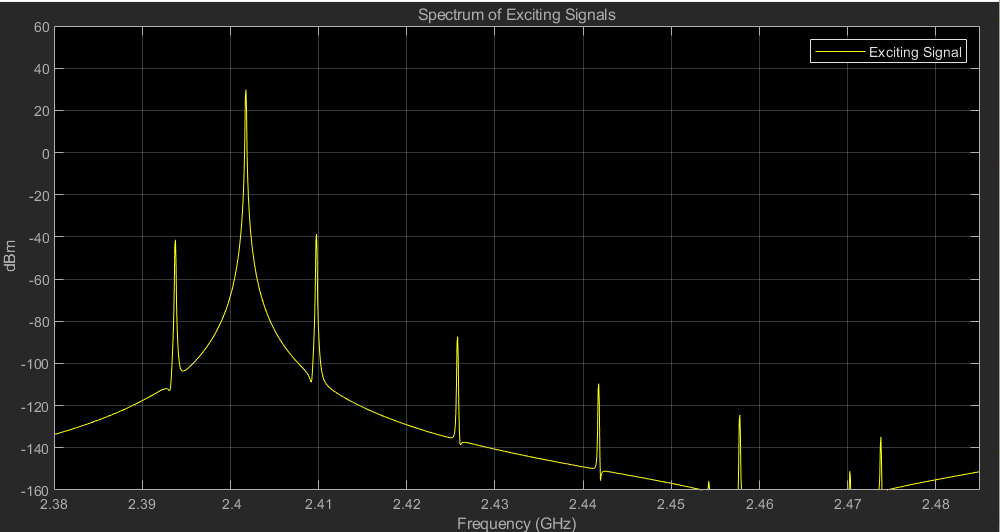}
    \caption{The spectrogram of exciting signal}
    \label{fig:7}
\end{figure}

\begin{figure}[ht]
    \centering
    \includegraphics[width=0.5\textwidth]{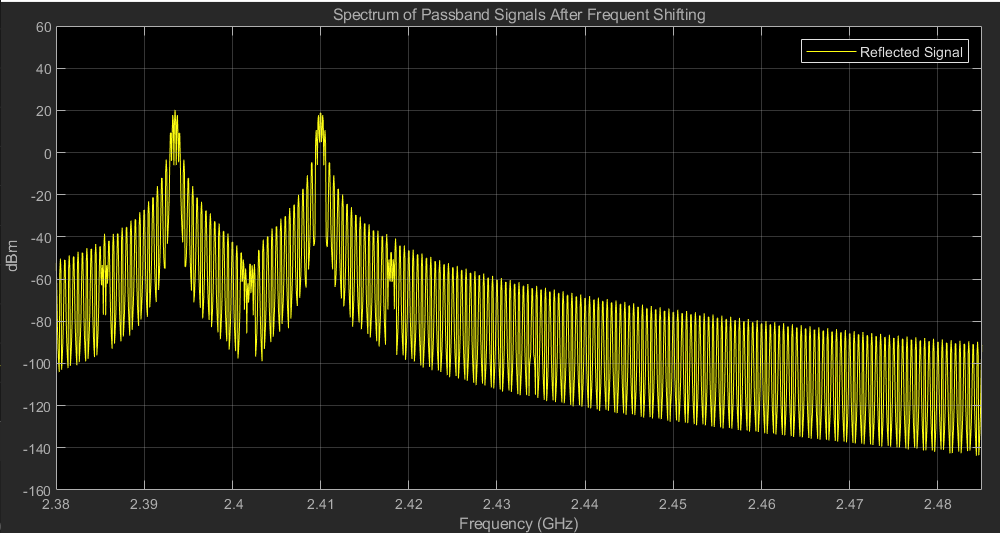}
    \caption{The spectrogram of reflected signal}
    \label{fig:8}
\end{figure}

\subsection{BER for different tag data}
There is a strange phenomenon in our experiment that different types of data will lead to completely different BER results. This experiment is done without awgn and uses GMSK modulation. The size of the tag data is 200 bits. The expected BER should be around 0. The result is shown in Figure 10.
\begin{figure}[ht]
    \centering
    \caption{BER for different tag data}
    \begin{tabular}{|c|c|}
        \hline
        data type & BER \\
        \hline
        random 0-1 string  & 0.257 \\
        \hline
        all 1 string & 0 \\
        \hline
        all 0 string & 0 \\
        \hline
        random string generated by  & 0 \\
        the regular expression $(00|11)^{*}$ & \\ 
        \hline
    \end{tabular}
\end{figure}

The reason for it is analyzed as follows:

\textbf{Frequency deviation:} The frequency deviation is implemented in Matlab's Bluetooth Toolbox. As we don't explore deep into the concrete algorithm of that, we don't know the exact number of it. We tried to fix it by changing the frequency deviation we use but the result only got worse.

\textbf{Modulation methods:} The modulation method of Bluetooth in the real world is GFSK and in Matlab it is GMSK. Though we try to replace it by FSK but still have the same problem. 

\textbf{The effect of Dsp.DigitalUpConverter:} In the next part, we find that using the DigitalUpConverter will increase the BER while other settings are the same. So it may also have affect on the accuracy of translating signals.

\subsection{BER comparison tests}
We choose random strings generated by the regular expression $(00|11)^{*}$ to bypass the problem above. Other settings are the same as in Part D. We change the Eb/No of the awgn channel to see the impact of it. We do each part of the experiment 10 times to reduce randomness. Figures 11–14 show the BER/dBNo results of different settings. Figure 11 is the result of an end-to-end model of RBLE with awgn channel and it uses GMSK as the modulation method. Figure 12 is much the same but uses FSK modulation. Figure 13 is a BLE model that simply transfers messages through awgn channel. For the experiments above, we add awgn to the intermediate band of the signal. We also have an experiment with a BLE model whose awgn is added to the baseband part. The result is shown in Figure 14.

By comparing Figures 11 and 12, we observe that the performance of GMSK and FSK is similar. The BER is high when Eb/No is low. As Eb/No increases, the BER decreases. When Eb/No is around 20, the BER of both methods goes down to 0.

By comparing Figures 11 and 13, we see the influence of the direct frequency shift on the BER/EbNo curve. That represents the robustness and reliability of RBLE. If there is no direct frequency shift, the BER will go to 0 earlier, as Figure 13 shows.

By comparing Figures 13 and 14, we observe that transforming the baseband signal to the IF band will introduce huge errors in BER. The BER goes down to 0 when Eb/No equals 6 without this transformation while the BER goes down to 0 when Eb/No equals 16 using this transformation. The reason for these huge errors may be the Matlab DigitalUpConverter and DigitalDownConverter.

\begin{figure}[ht]
    \centering
    \includegraphics[width=0.5\textwidth]{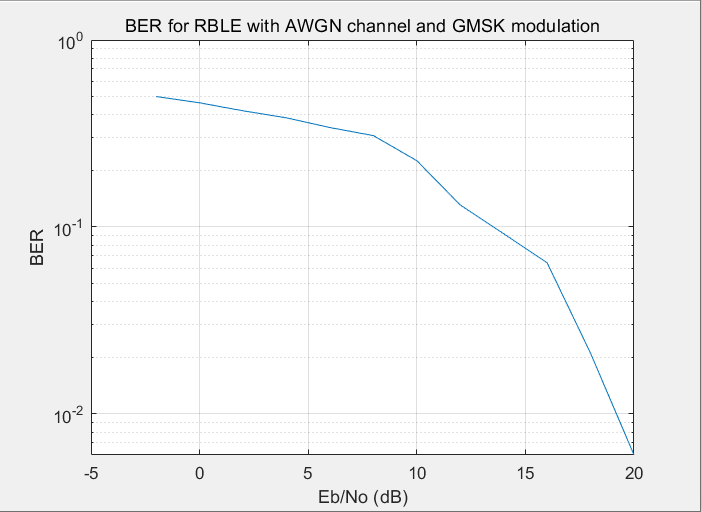}
    \caption{The BER/EbNo curve for RBLE with AWGN channel and GMSK modulation}
    \label{fig:9}
\end{figure}

\begin{figure}[ht]
    \centering
    \includegraphics[width=0.5\textwidth]{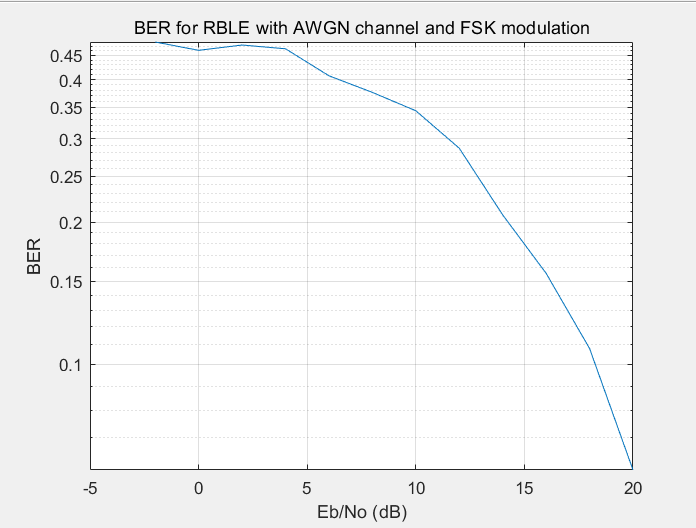}
    \caption{The BER/EbNo curve for RBLE with FSK channel and GMSK modulation}
    \label{fig:10}
\end{figure}

\begin{figure}[ht]
    \centering
    \includegraphics[width=0.5\textwidth]{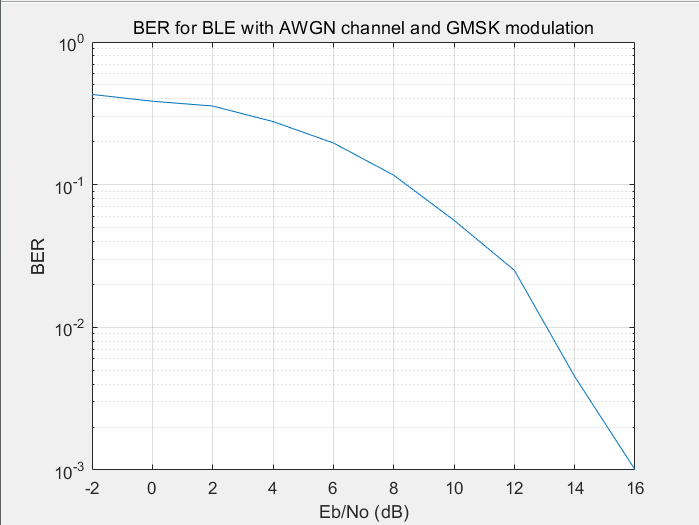}
    \caption{The BER/EbNo curve for BLE with AWGN channel and GMSK modulation}
    \label{fig:11}
\end{figure}

\begin{figure}[ht]
    \centering
    \includegraphics[width=0.5\textwidth]{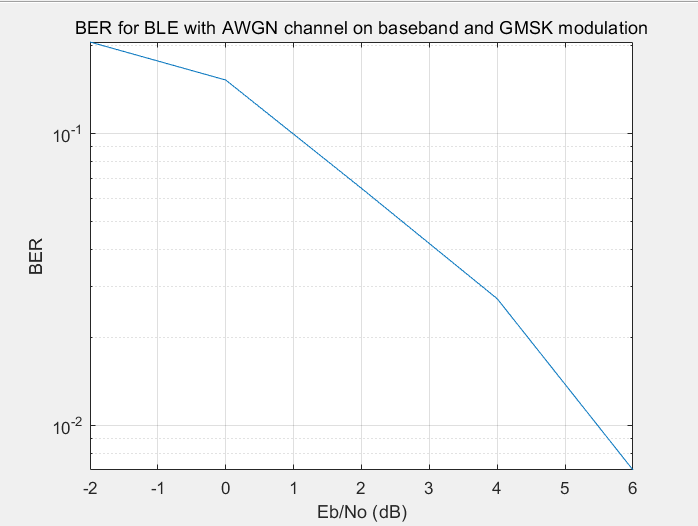}
    \caption{The BER/EbNo curve for BLE with AWGN channel and GMSK modulation}
    \label{fig:12}
\end{figure}

\section{Outlook adn Future Work}
In this part, we will have a discussion about the situations that are suitable for RBLE abd the future work of our experiment.

The first scenario is in the logistics link. For example, if a large number of goods need to be registered, we can solve it by pasting RBLE tags on the goods. Compared to active solutions such as scanning the barcode information on the goods to register, RBLE is more convenient and faster. Compared with other backscatter systems, RBLE does not require additional receivers, is more robust and can guarantee a lower error rate. However, the current problem is that the cost of RBLE tag is not as low as barcode paper.

The second scenario is in the home. With the popularity of smart home, there are more and more smart devices in each home. If all smart home devices use the RBLE scheme, it means that users do not need to care about the replacement of battery, and at the same time people can use the phone to control all the devices.

The backscatter technique should also have more possibilities if it can solve the problem of the current need for a Bluetooth transmitter to transmit a single frequency signal. For example, RBLE works completely on commodity BLE. If we can reflect arbitrary Bluetooth signals and add tag information, we can use serial backscatter to solve the current problem of distance transmission. In smart home use, we also do not need a Bluetooth transmitter. Backscatter devices only need to reflect the signal from the active Bluetooth device in the home.

As for our RBLE simulation experiment, I think there is more work that can be done in the future as follows:

\textbf{Encodings for the information:} In order to enhance the reliability of data transmission, we can encode the original data in different ways. Using simulation experiments helps us to choose an appropriate encoding.

\textbf{Test BER through different noise channels:} There are different types of loss models for simulating the process of transferring signals. We choose the simplest one in our experiment but we can replace it with others to get more results.

\textbf{Support different modes of Bluetooth 5.x environment:} The Bluetooth toolbox implements all four modes of Bluetooth 5.0. We only implement RBLE in the 1Mbps mode. Future work can be done to support all the modes in Bluetooth 5.0 and it will have a wider usage.

\section{Conclusion}
In this paper, we have implemented the Matlab simulation of RBLE, which is a reliable BLE backscatter system that works with a single commodity receiver. We analyze the reason for the BER is mainly because of the conversion progress between the baseband signal and the IF band signal. Future work includes implementing GFSK modulation in the RBLE to be closer to the real case, changing BLE packet structures in order to support different modes of Bluetooth 5.x environments.

\vspace{12pt}

\end{document}